%% file: main.tex
\documentclass[10pt, conference]{IEEEtran}
\usepackage{tcolorbox}
\usepackage{fbox}
\usepackage{xcolor}
\usepackage{booktabs}
\usepackage{verbatim}
\usepackage{pgfplots}
\usepackage{filecontents}
\usepackage{tikz}
\usepackage{subfigure}
\usepackage{graphics}
\usepackage{multirow}
\usepackage{fancybox}
\usepackage[switch]{lineno}
\usepackage{balance}
\usepackage[flushleft]{threeparttable}
\usetikzlibrary{patterns}
\usepackage{hhline}
\usepackage{layouts}
\usepackage{arydshln}
\usepackage[linesnumbered,ruled,vlined]{algorithm2e}
\usepackage{amssymb}
\usepackage{amsmath}
\usepackage{stackengine}
\usepackage{setspace}
\usepackage{microtype}
\usepackage{mathtools}
\usepackage{soul}

\usepackage[hang, flushmargin]{footmisc}
\usepackage[colorlinks, linkcolor=blue, anchorcolor=blue, urlcolor=blue, citecolor=blue]{hyperref}
\usepackage{footnotebackref}

\SetCommentSty{mycommfont}

\pgfplotsset{compat=1.16}

\definecolor{red}{rgb}{1, 0, 0}
\definecolor{green}{rgb}{0, 1, 0}
\definecolor{blue}{rgb}{0, 0, 1}
\definecolor{dark-green}{rgb}{0, 0.6902, 0.31373}

\newcommand{\rv}[1]{#1}

\newcommand{\logtext}[1]{``\textls[-70]{\texttt{#1}}"}

\newcommand\tool{LogPPT }
\newcommand{\repourl}{\url{https://github.com/LogIntelligence/LogPPT}}

\def\BibTeX{{\rm B\kern-.05em{\sc i\kern-.025em b}\kern-.08em
    T\kern-.1667em\lower.7ex\hbox{E}\kern-.125emX}}
\begin{document}

\title{Log Parsing with Prompt-based Few-shot Learning}

\author{\IEEEauthorblockN{
Van-Hoang Le$^{1}$ and
Hongyu Zhang$^{2}$\IEEEauthorrefmark{2}\thanks{\IEEEauthorrefmark{2}Hongyu Zhang is the corresponding author.}
\IEEEauthorblockA{$^1$School of Information and Physical Sciences, The University of Newcastle, Australia}
\IEEEauthorblockA{$^2$School of Big Data and Software Engineering, Chongqing University, China}
\IEEEauthorblockA{vanhoang.le@uon.edu.au, hyzhang@cqu.edu.cn}
}\\\vspace{-15pt}}

\maketitle

\begin{abstract}
Logs generated by large-scale software systems provide crucial information for engineers to understand the system status and diagnose problems of the systems. Log parsing, which converts raw log messages into structured data, is the first step to enabling automated log analytics. Existing log parsers extract the common part as log templates using statistical features. However, these log parsers often fail to identify the correct templates and parameters because: 1) they often overlook the semantic meaning of log messages, and 2) they require domain-specific knowledge for different log datasets. To address the limitations of existing methods, in this paper, we propose \tool to capture the patterns of templates using prompt-based few-shot learning. \tool utilises a novel prompt tuning method to recognise keywords and parameters based on a few labelled log data. In addition, an adaptive random sampling algorithm is designed to select a small yet diverse training set. We have conducted extensive experiments on 16 public log datasets. The experimental results show that \tool is effective and efficient for log parsing.
\end{abstract}

\begin{IEEEkeywords}
log parsing, few-shot learning, prompt-tuning, deep learning
\end{IEEEkeywords}

\section{Introduction}
\input{sections/introduction}

\section{Background and Motivation}
\input{sections/background.tex}

\section{Approach}
\input{sections/approach.tex}

\section{Experimental Design}
\input{sections/ex_design.tex}

\section{Experimental Results}
\input{sections/ex_results.tex}

\section{Discussion}
\input{sections/discussion.tex}

\section{Related Work}
\input{sections/related_work.tex}

\section{Conclusion}
\input{sections/conclusion.tex}

\section*{Acknowledgment}
This work is supported by Australian Research Council (ARC) Discovery Projects (DP200102940, DP220103044).
We also thank anonymous reviewers for their insightful and constructive comments, which significantly improve this paper.


\bibliographystyle{IEEEtran}
\bibliography{ref}
\balance

\end{document}

%% file: sections/introduction.tex
Large-scale software-intensive systems often produce a large volume of logs to record runtime status and events for troubleshooting purposes.
Logs play an important role in the maintenance and operation of software systems, which allow engineers to better understand the system's behaviours and diagnose problems. The rich information included in log data enables a variety of log analytics tasks, such as anomaly detection~\cite{du2017deeplog, zhang2019robust, zhang2020anomaly, zhang2023semi}, root cause analysis~\cite{lu2017log, gurumdimma2016crude}, failure prediction~\cite{das2018desh, zhang2018prefix}, and log compression~\cite{liu2019logzip, wei2021feasibility}. Among them, the first and foremost step is log parsing, which parses free-text raw log messages into a structured format~\cite{zhu2019tools}. The structured log data from log parsing are fed to various machine learning (ML) or deep learning (DL) models to perform many downstream analysis tasks.

\begin{figure}[h]
    \centering
    \includegraphics[width=.98\linewidth]{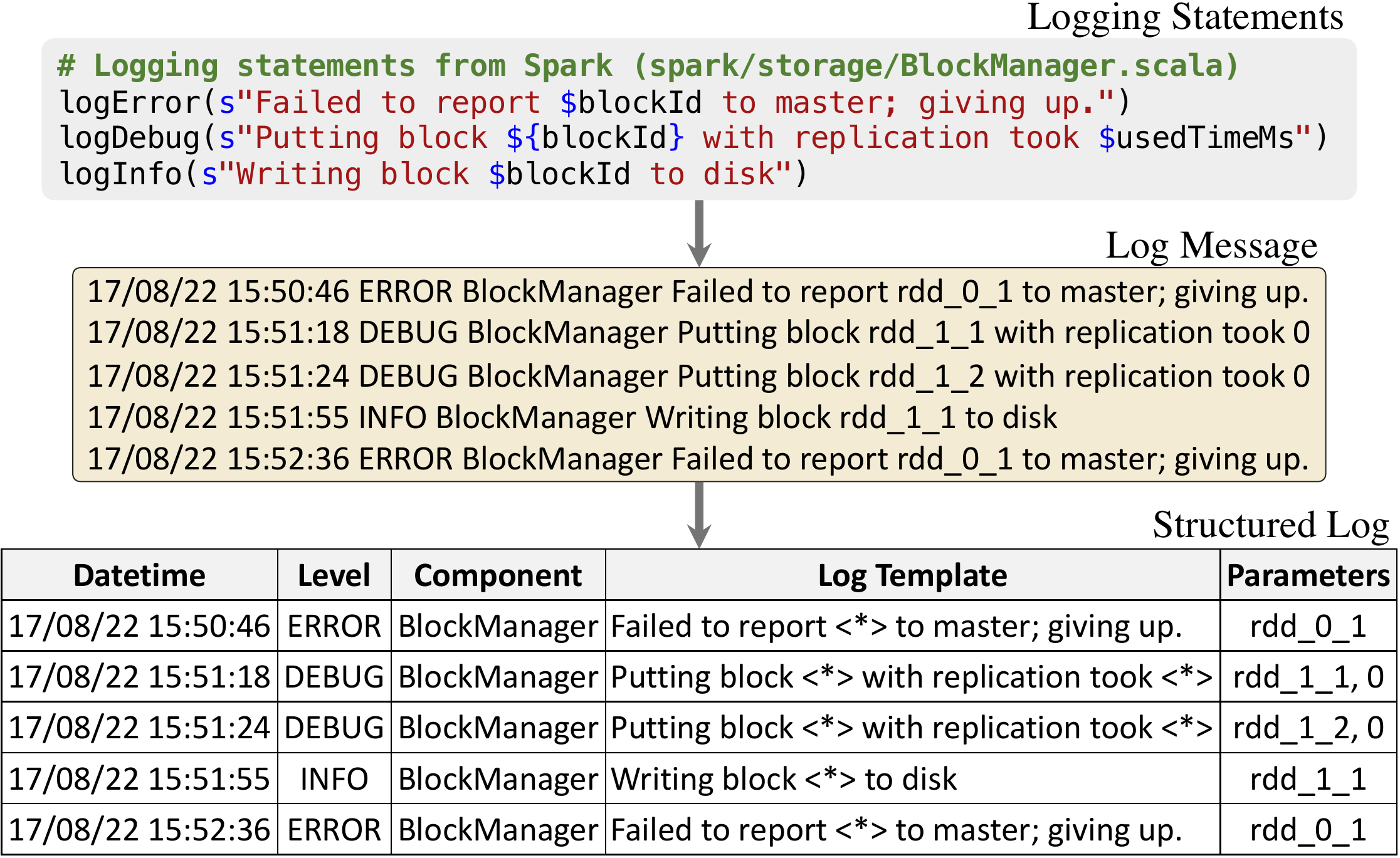}
    \caption{An example of log parsing from Spark}
    \label{fig:log-parsing-example}
\end{figure}

Log parsing is the task of converting a raw log message into a specific log template. 
As shown in Figure~\ref{fig:log-parsing-example}, log messages are generated from logging statements in the source code. A log message usually contains a header that is automatically produced by the logging framework and includes information such as component and verbosity level. The log message body (log message for short) typically consists of two parts: 1) \textit{Template} - constant strings (or keywords) describing the system events; 2) \textit{Parameters} - dynamic variables, which vary during runtime and reflect system runtime information. For example, in the first log message in Figure~\ref{fig:log-parsing-example}, the header (i.e., \logtext{17/08/22 15:50:46}, \logtext{INFO}, and \logtext{BlockManager}) can be easily distinguished through regular expressions. The log message consists of a template \logtext{Putting block <*> with replication took <*>} and the parameters including \logtext{rdd\_1\_1} and \logtext{0}.

To achieve automated log parsing, many data-driven approaches~\cite{he2017drain, du2016spell, dai2020logram, nagappan2010abstracting} have been proposed over the years to extract the common parts that constantly occur among log messages as templates and the dynamic parts that vary during runtime as parameters. Although making progress, existing log parsers still suffer from unsatisfactory accuracy, which may significantly affect the follow-up analysis such as log-based anomaly detection~\cite{le2021log}. For example, the existing state-of-the-art log parsers Drain~\cite{he2017drain} and AEL~\cite{jiang2008abstracting} only achieve an average Parsing Accuracy of 0.34 and 0.28
on 16 log datasets~\cite{khan2022guidelines}.
We have identified the following limitations of the existing log parsers:
\begin{itemize}
    \item \textbf{Accuracy:} Existing log parsers extract common parts as templates using statistical features (e.g., word length, log length, frequency) and ignore the semantic meaning of log messages. Without considering the semantic information, traditional log parsers tend to misidentify parameters as keywords~\cite{liu2022uniparser} in many cases (e.g., when encountering previously unseen log templates).
    \item \textbf{Robustness:} Existing log parsers are not robust across different types of logs because they require domain-specific knowledge for different datasets~\cite{nedelkoski2020self-parsing}.
    The domain-specific knowledge includes data pre-processing (e.g., defining regular expressions) and hyper-parameter settings (e.g., the number of clusters or similarity threshold).
    The accuracy of these log parsers could be significantly affected by the input domain-specific knowledge.
    For example, without the pre-processing step, the parsing accuracy can decline by 6.1\%-73.5\%~\cite{he2016evaluation}.
    When applying the existing log parsers to a new log dataset, due to different logging formats and behaviours, time-consuming adjustment of hyper-parameters and 
    regular expressions are needed~\cite{liu2022uniparser}.
 
\end{itemize}


To overcome the above-mentioned limitations, in this paper, we propose LogPPT, a novel log parser with prompt-based few-shot learning. LogPPT is able to capture the semantic information of log messages to identify keywords and parameters in log messages by learning from only a few labelled log messages. First, we design an Adaptive Random Sampling algorithm that can sample a small and diverse set of log messages to label as the training data.
The training data is a set of labelled logs that contain raw log messages and the corresponding ground truth templates.
Second, to effectively train a model with a few labelled log data, we tune a pre-trained language model (e.g., RoBERTa~\cite{liu2019roberta}) to predict a specific virtual label token \rv{(\logtext{PARAM}, an acronym for parameters)} at the position of parameters in the log message in a few-shot learning manner. The \rv{embedding vector} for the virtual label token \logtext{PARAM} is generated based on the word distribution from language model predictions and the unlabelled log dataset. After training, \tool can be directly applied to parse new log data. Our proposed method does not require any pre-processing step and uses the same set of hyper-parameter values for different datasets, which is robust across different logging formats and behaviours, and more generalised than existing approaches.

We have evaluated \tool on 16 public log datasets~\cite{zhu2019tools}. \tool achieves over 0.9 average Group Accuracy~\cite{zhu2019tools} and Parsing Accuracy~\cite{khan2022guidelines, liu2022uniparser} when using only 32 labelled samples. The experimental results show that \tool is effective and efficient. It outperforms state-of-the-art parsers by 16\% on Group Accuracy~\cite{zhu2019tools} and about 84\% on Parsing Accuracy~\cite{liu2022uniparser}. Moreover, \tool is also robust across different log datasets. 

To summarise, our main contributions are as follows:
\begin{itemize}
    \item We propose LogPPT, a prompt-based few-shot log parser that can precisely capture the patterns of templates and parameters in log messages. \tool uses a novel prompt tuning method to effectively learn the semantic information from a few labelled log samples. The proposed approach does not require manually-defined regular expressions for pre-processing and uses the same set of hyper-parameter values for every dataset, thus can quickly adapt to new log datasets. 
    \item We evaluate \tool on 16 public log datasets, and the results demonstrate that \tool outperforms existing approaches. The experimental results confirm the effectiveness and efficiency of our proposed method.
\end{itemize}

%% file: sections/background.tex
\subsection{Log Parsing}
Log parsing is one of the first steps for log analysis tasks~\cite{zhu2019tools}. It is a process to extract the static log template parts and the corresponding dynamic parameters (or variables) from free-text raw log messages. For example, Figure~\ref{fig:log-parsing-example} shows an example of logs of the Spark system, where Datetime, Component, and Level fields are the log header generated by the logging framework and are generally easy to extract. 
The log template \logtext{Putting block <*> with
replication took <*>} associated with parameters (e.g., \logtext{rdd\_1\_1}, \logtext{0}), in contrast, \rv{is} often difficult to identify. The goal of log parsing is to convert each log message into a specific log template and extract the corresponding parameters~\cite{zhu2019tools, liu2022uniparser}.


\rv{The straightforward way of log parsing relies on handcrafted regular expressions or grok patterns to extract log templates and parameters~\cite{zhu2019tools}.
However, manually writing 
regular expressions to parse a huge volume of logs is time-consuming and error-prone~\cite{zhu2019tools}. 
Some studies~\cite{xu2009detecting, nagappan2009efficiently} extract the log templates from logging statements in the source code to compose regular expressions for log parsing. However, it is not applicable in practice since the source code is often unavailable, especially for third-party libraries~\cite{zhu2019tools}. Therefore, regular expression matching 
often serves as a pre-processing step to (1) separate headers and content (which contains log templates and dynamic parameters) from raw log messages, and (2) abstract some special information such as IP address and ID 
to improve parsing accuracy.}
To achieve the goal of automated log parsing, many data-driven approaches have been proposed to identify log templates as the frequent part of log messages.
Data-driven log parsing approaches can be divided into three main groups:

{1) Frequent pattern mining.}
Some approaches, including SLCT~\cite{vaarandi2003data}, LFA~\cite{nagappan2010abstracting}, and Logram~\cite{dai2020logram}, find frequent patterns which emerge constantly across the entire log dataset. They leverage the token position or $n$-gram information to extract log templates based on frequent pattern mining.

{2) Similarity-based clustering.}
These approaches apply various clustering algorithms to group similar logs and consider logs under the same group belonging to the same template. Representative methods include LKE~\cite{fu2009execution}, LogSig~\cite{tang2011logsig}, and LenMa~\cite{shima2016length}, which compute distances between two log messages or their signature to cluster them based on similarity.

{3) Heuristics-based parsing.}
AEL~\cite{jiang2008abstracting}, Spell~\cite{du2016spell}, or Drain~\cite{he2017drain} propose heuristics-based log parsing methods that leverage unique characteristics from log messages to extract common templates efficiently.

Although making progress, traditional log parsers are still criticized for unsatisfactory parsing accuracy due to the omission of semantic information or improper evaluation metrics. Recent studies~\cite{khan2022guidelines, liu2022uniparser} show that traditional approaches focus more on grouping logs and fail to identify the correct templates and parameters.
For example, in Figure~\ref{fig:log-parsing-example}, some tokens (such as \logtext{rdd\_0\_1} and \logtext{0}) are identified as keywords by traditional log parsers because they do not vary in different log messages. However, these tokens should be classified as parameters considering their semantic meanings.
Besides, existing log parsers are not robust across different log datasets.
They require domain-specific knowledge to define regular expressions for pre-processing of different log data~\cite{nedelkoski2020self-parsing}.
For example, on the HDFS dataset~\cite{he2016evaluation, hdfsdataset2022}, $block\_id$ (e.g., \logtext{blk\_-6670958622368987959}) information is abstracted from logs by using a regular expression \logtext{blk\_-?$\backslash$d+}. 
For a new dataset such as BGL~\cite{he2016evaluation, bgldataset2022}, this regular expression must be changed to match the $core\_id$ such as \logtext{core.2275} (i.e., \logtext{\textbf{blk\_}-?$\backslash$d+} $\rightarrow$ \logtext{\textbf{core.}$\backslash$d+}).
Moreover, existing log parsers require specific hyper-parameters (e.g., number of clusters or similarity threshold) for different datasets to optimize the performance. For example, Drain~\cite{he2017drain} uses a low \textit{similarity threshold} of 0.2 for the HealthApp dataset and a high \textit{threshold} of 0.6 for the Proxifier dataset~\cite{zhu2019tools}.
Due to different logging formats and behaviours, when facing a new log dataset, existing log parsers have to adjust the hyper-parameters and reconfigure the regular expressions for pre-processing~\cite{liu2022uniparser}. 


\subsection{Language Models}
\label{sec:language-models}
\subsubsection{Pre-training and Fine-tuning}
Pre-trained models have been shown effective in many natural language processing (NLP) tasks. These language models (LM), such as BERT~\cite{devlin2018bert} and T5~\cite{raffel2020T5}, are generally pre-trained using the Masked Language Modelling (MLM) objective. During the pre-training phase, the model learns to predict randomly masked input tokens.
Based on the idea that log is actually a natural language sequence~\cite{zhang2019robust}, some studies~\cite{le2021log, li2020_swisslog, tao2021logstamp} have leveraged pre-trained language models such as BERT~\cite{devlin2018bert} to analyse log data. Language models are pre-trained on large-scale unlabelled corpus and then fine-tuned to perform downstream tasks.

\begin{figure}[h]
    \centering
    \includegraphics[width=.95\linewidth]{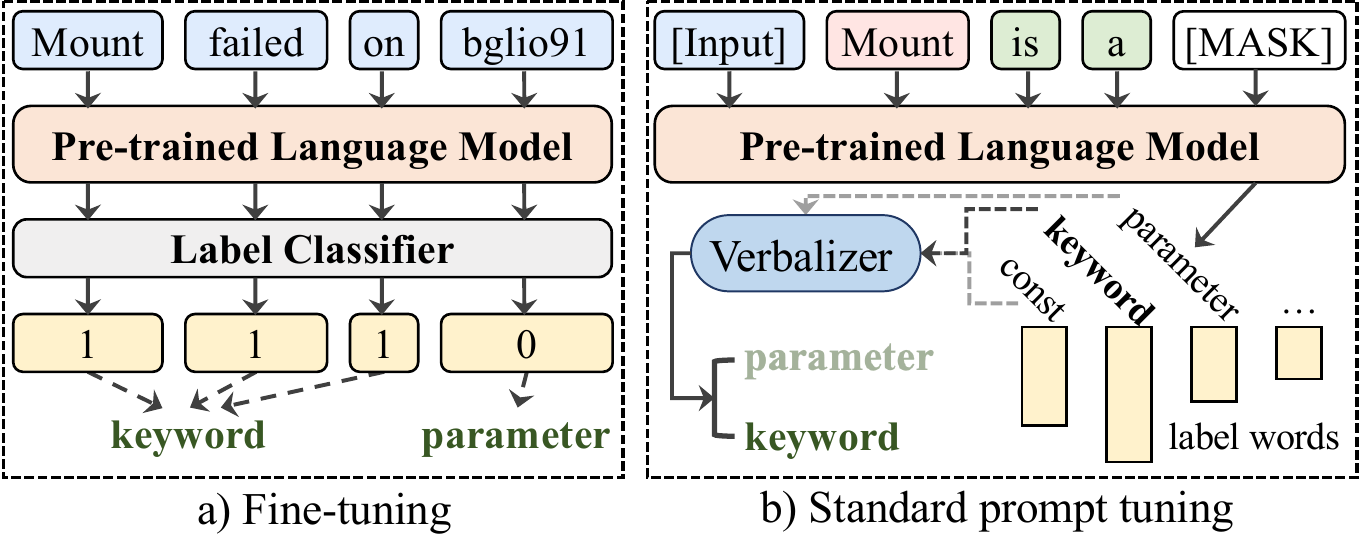}
    \caption{An illustration of fine-tuning and prompt tuning for log parsing}
    \label{fig:pretrained-background}
\end{figure}

\textbf{Fine-tuning} a pre-trained model for downstream tasks~\cite{devlin2018bert, radford2018improving} is a prevalent paradigm in the NLP field that further trains the model in a supervised way. As shown in Figure~\ref{fig:pretrained-background}(a), a straightforward way to apply fine-tuning for log parsing is to convert the log parsing task into the token classification problem. The model can easily extract keywords and form log templates by classifying whether a token in log messages is a keyword or parameter (binary classification) using an additional classifier. However, the inconsistency between pre-training objectives and the fine-tuning objective (i.e., classification) restrains the use of rich knowledge distributed in pre-trained models~\cite{wang2022no, han2021ptr}, leading to sub-optimal results. Besides, the performance of fine-tuning significantly depends on the scale of downstream data.

\subsubsection{Prompt Tuning}
Recently, prompt tuning~\cite{han2021ptr, gao2021making, ma-etal-2022-template, wang2022instructionner} has been proposed to close the gap between pre-training and downstream tasks.
Figure~\ref{fig:pretrained-background}(b) illustrates the concept of prompt tuning. Instead of designing a new training objective for each downstream task, prompt tuning rewrites the input by adding a natural language instruction such as ``[S] is a [MASK]" to reuse the masking objective for downstream tasks. Formally, standard prompt tuning employs a prompt template $T_{prompt}(.)$ to convert the input $X$ to prompt input $X_{prompt} = T_{prompt}(X)$. The prompt template is a textual string with unfilled slots to fill the input $X$ and a label slot [MASK]. 

For log parsing, a standard prompt template consists of three unfilled slots to fill the input log message, the token needed to be identified, and the label for the processing token. For example, in Figure~\ref{fig:pretrained-background}(b), the prompt template is in the form of ``[X] [S] is a [MASK]", where [X], [S], and [MASK] are the unfilled slots for the input log message, token, and label, respectively. The LMs then try to fill the label slot [MASK] with label words such as \textit{keyword} or \textit{variable}.
After that, a verbalizer is used to map each predicted label word to a class for the input token. In Figure~\ref{fig:pretrained-background}(b), the verbalizer contains label words sets of ``[\textit{const, keyword}]" for keywords and ``[\textit{parameter}]" for parameters. By enumerating over all tokens in a log message, we can extract the corresponding template and parameters.

According to the flexibility of the prompt template, standard prompt tuning techniques can be categorized into two types: hard prompt and soft prompt. We briefly introduce each prompt type in the following.

\textbf{Hard Prompt.} Hard prompt or discrete prompt~\cite{han2021ptr, gao2021making} is a technique that modifies the input by adding fixed natural language instructions. Hard prompt templates usually correspond to natural language phrases~\cite{liu2021pre}, in which each token in prompt templates is meaningful and understandable. Although hard prompt has shown promising performance~\cite{gao2021making}, the template design and the label word choices are challenging because it requires task-specific knowledge.

\textbf{Soft Prompt.} Soft prompt~\cite{li2021prefix, qin-eisner-2021-learning} is an alternative to hard prompt. Instead of using fixed discrete words as in hard prompt, soft prompt uses \textit{virtual tokens}, which are in the form of continuous vectors and can be learnt during the tuning stage, to construct prompt templates. The soft prompt is proposed to remove the constraints of manually selecting a prompt template in the hard prompt.

Although achieving promising results in various NLP tasks, standard prompt tuning is insufficient for the log parsing task because (1) it needs to enumerate all span candidates, which is inelegant and time-consuming~\cite{wang2022instructionner}, and (2) it is sensitive to noises 
(see Section~\ref{sec:diss-why-logppt-work} for details).

In this paper, we apply prompt tuning to achieve the goal of log parsing with a few labelled training data. However, instead of using standard prompt tuning, we leverage the paradigm of template-free prompt~\cite{ma-etal-2022-template} for log parsing, which does not require prompt templates as the instruction. In template-free prompt~\cite{ma-etal-2022-template}, an additional \textit{virtual label token} is generated and plays the role of prompt instructions as in standard prompt tuning. Then, the model learns to predict the \textit{virtual label token} at the positions of parameters and the original token at the positions of keywords using a custom MLM objective. The template-free prompt tuning method~\cite{ma-etal-2022-template} addresses \rv{the} major limitations of standard prompt 
by (1) relaxing the burden of manually selecting prompt templates~\cite{ma-etal-2022-template} and (2) performing one-pass decoding to process all tokens 
simultaneously, which is more efficient compared to the time-consuming enumeration process of standard prompts~\cite{ma-etal-2022-template, wang2022instructionner}.

%% file: sections/approach.tex
In this section, we describe the proposed \tool approach. To overcome the limitations of existing approaches, we train a model to capture the patterns of templates and parameters based on the context information of log messages using rich knowledge derived from language models pre-trained on large corpora. Specifically, we apply the paradigm of prompt tuning~\cite{ma-etal-2022-template} to enable few-shot log parsing to better transfer the knowledge from pre-trained language models to log parsing.
To make the best use of prompt tuning, it is essential to select an optimized labelled training set for our method. Therefore, we introduce an Adaptive Random Sampling algorithm to effectively select a small number of samples 
for training. 

\begin{figure}[htbp]
    \centering
    \includegraphics[width=.98\linewidth]{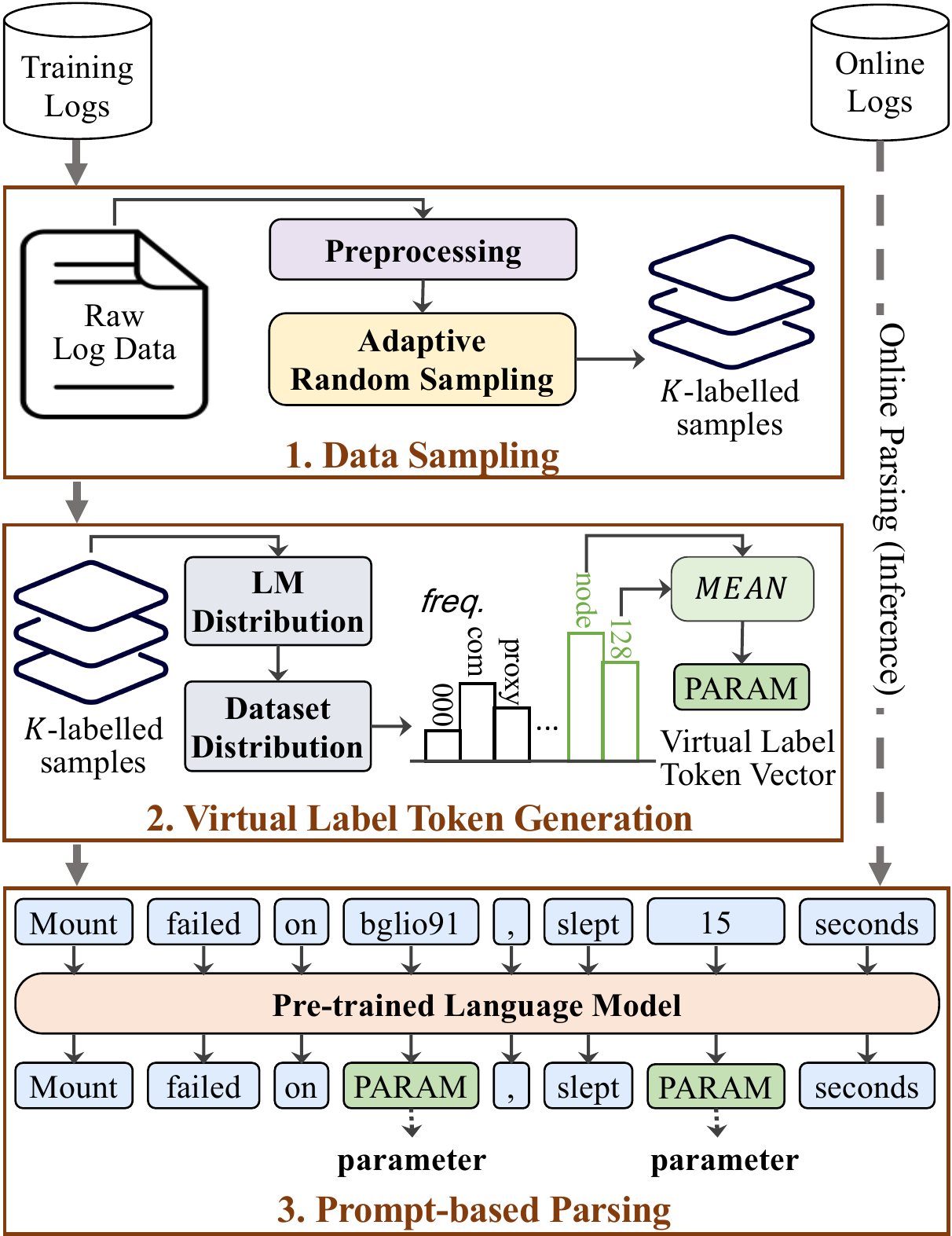}
    \caption{An overview of LogPPT}
    \label{fig:overview}
\end{figure}

The overview of the proposed approach is shown in Figure~\ref{fig:overview}.
In the following, we first present the problem formulation in Section~\ref{sec:prob-define}. Then, we describe the few-shot data sampling method in Section~\ref{sec:data-sampling}. Section~\ref{sec:prompt-tuning-log-parsing} describes the training process, which consists of three modules, including a pre-train\rv{ed} language model, a virtual label token generation module, and a training objective. Finally, we describe how to apply \tool for online parsing in Section~\ref{sec:online-parsing}.

\subsection{Problem Definition}
\label{sec:prob-define}
In this work, we transform the log parsing task into a parameter recognition problem where only a small number of labelled examples are used for training by adopting a novel prompt tuning method~\cite{ma-etal-2022-template}. 
Specifically, 
for a new dataset $\mathcal{D}$, we tune a pre-trained language model, $\mathcal{M}$, to recognise keywords and parameters in a log message through prompt tuning. The model takes the input of a raw log message consisting of $n$ tokens, $X = \{ x_1, x_2, \dots, x_n\}$ and predicts a virtual label token \logtext{PARAM} at the position of parameters. For keywords, the model remains to predict the original tokens.
Formally, the model $\mathcal{M}$ is trained to generate the output, $Y=\{y_1, y_2, \dots, y_n\}$, where:

\begin{equation}
\label{eq:prompt-function}
y_i = \mathcal{M}(x_i) =
\begin{cases}
    \logtext{PARAM} & \text{if } x_i \text{ is a parameter}\\
    x_i & \text{if } x_i \text{ is a keyword}
\end{cases}
\end{equation}

For example, as shown in Figure~\ref{fig:overview}, the model is trained to predict the parameter \logtext{blgio91} as a label token \logtext{PARAM}. For keywords such as \logtext{failed}, the model will predict the original words. \logtext{PARAM} is a specific virtual token that does not have any linguistic meaning. It 
indicates parameters in log messages and guides the model to recognise those parameters based on their relations with the \logtext{PARAM} token.
The embedding vector of \logtext{PARAM} is calculated based on the most frequent parameters in log messages.
\logtext{PARAM}, therefore, is generated using both labelled training data and unlabelled data to better represent the meaning of parameters in log messages. In the online parsing (inference) phase, all tokens with $y_i = \logtext{PARAM}$ are considered parameters, and other tokens are included in the log template.



\subsection{Few-shot Data Sampling}
\label{sec:data-sampling}
During the training phase, our proposed method requires a small amount of labelled log data as the training dataset. 
To collect accurately labelled samples with low manual effort, we propose a simple yet effective approach to select a small number ($K$) of labelled samples.
Firstly, training log messages are cleaned by applying some commonly-used pre-processing techniques~\cite{zhang2019robust, le2021log}, such as removing all non-character tokens, stop words or camel case.
Then, we propose to use an Adaptive Random Sampling algorithm from Adaptive Random Testing~\cite{chen2004adaptive} to obtain a diverse and evenly distributed sample set. Algorithm~\ref{alg:data-sampling} describes the adaptive random sampling based algorithm for few-shot data selection.

\begin{algorithm}
    \small
    \SetAlgoLined
    \DontPrintSemicolon
    \caption{Few-shot Data Sampling}
    \label{alg:data-sampling}
    \KwData{
    $\mathcal{D}$: Log dataset \\
    \hspace*{3.1em}$K$: The number of collected samples}
    \KwResult{$\mathcal{D}_{train}$: a set of $K$-labelled samples}
    $\widehat{L} \gets $ pre-process($\mathcal{D}$)
    \tcp*[f]{$\widehat{L}_i = \{cln, org\}$: clean and original logs}\\
    $\mathcal{D}_{train} \gets \O$ \tcp*[f]{initialize the training set} \\
    $\mathcal{S} \gets \{ l~|~l \in \widehat{L} \textbf{ and } l.cln \text{ is the shortest cleaned log} \}$ \\
    \While {$K > 1$} {
        $\widehat{C} \gets \O$ \tcp*[f]{initialize candidate set} \\
        \For(\tcp*[f]{$\eta = 32$}){$i = 1 \to \eta$} {
            $\widehat{C}.\textbf{add}(\{ \text{random } c \in \widehat{L}|~c.cln \notin \widehat{C} \And c.org\notin\mathcal{S}\}) $ \\
        }
        
        \tcc{compute the similarities between logs in $\widehat{C}$ and $\mathcal{S}$}
        $\Delta \gets \O$ \\
        \For {$c=\{cln, org\} \in \widehat{C}$} {
            $\delta \gets 0$ \\
            \tcc{find the nearest neighbour of $c$ in $\mathcal{S}$ and calculate the similarity between $c$ and its nearest neighbour}
            \ForEach {$l=\{cln, org\} \in \mathcal{S}$} {
                $\delta = \textbf{MAX}(\delta, \textbf{similarity}(c.cln, l.cln))$ \\
            }
            $\Delta.\textbf{add}(\delta)$ \\
        }
        \tcc{select the candidate with the longest distance/smallest similarity to its nearest neighbour in $\mathcal{S}$}
        $\mathcal{S}.\textbf{add}(\{c \in \widehat{C}~|~\Delta_c \text{ is smallest}\})$ \\
        $K \gets K - 1$ \\
    }
    \tcc{label the sample set $\mathcal{S}$ of $K$ samples as the training set}
    \ForEach{$s=\{ cln, orig\} \in \mathcal{S}$} {
        $\mathcal{D}_{train}$.\textbf{add}(\{$s.orig, template(s.orig)$\}) \\
    }
    \KwRet{$\mathcal{D}_{train}$}
\end{algorithm}

Algorithm~\ref{alg:data-sampling} takes a raw log dataset $\mathcal{D}$ and a desired number of samples in training set $K$. At line 1, all log messages in $\mathcal{D}$ are pre-processed by applying commonly-used pre-processing techniques~\cite{zhang2019robust, le2021log}. The result of this step is a set $\widehat{L} = \{ \dots, (cln, orig), \dots \}$ in which each element contains a clean log message and an original log message. At lines 2-3, the algorithm initializes the following two components: (1) an empty set $\mathcal{D}_{train}$, which is the result of the algorithm; \rv{(2) a set $\mathcal{S}$, which contains the shortest log message at first, to store selected log messages to label}. At lines 4-19, the algorithm iteratively selects \rv{one log message per iteration} based on their similarities until $\mathcal{S}$ contains $K$ samples. From lines 5-8, $\eta$ random candidate logs from $\widehat{L}$ are selected and stored in $\widehat{C}$. Then, for each candidate in $\widehat{C}$, the algorithm finds and calculates the similarity with its nearest neighbour in $S$ (lines 9-16). At line 17, the algorithm finds a candidate $c$ in $\widehat{C}$ which has the smallest similarity with its nearest neighbour (i.e., smallest $\Delta_c$) and inserts it to the sample set $\mathcal{S}$. The outer loop repeats until $\mathcal{S}$ contains $K$ elements. From lines 20-23, the algorithm collects the templates for all original log messages in $\mathcal{S}$ from user feedback and returns $\mathcal{D}_{train}$ as the final output.


\subsection{Prompt-Tuning for Log Parsing}
\label{sec:prompt-tuning-log-parsing}

In this work, we take advantage of prompt-tuning, which recently set the state-of-the-arts for many NLP tasks, by applying the entity-oriented LM objective~\cite{ma-etal-2022-template}. 
The essence behind this idea is that (1) most keywords in log statements are valid words and readable, which can be looked up in a dictionary~\cite{li2020_swisslog}, thus are easier to be predicted by the language model; and (2) parameters, in contrast, are constantly changing, which are hard to be predicted 
by the language model. 
In view of this, we transform the log parsing task into a label token prediction problem. Specifically, for parameters, we force the model to predict the virtual label token \logtext{PARAM}, while for keywords, the model is trained to predict the original words.

\subsubsection{Pre-trained Language Model}
Pre-trained language models~\cite{liu2019roberta, devlin2018bert, radford2019language, radford2018improving} have been shown to be effective in many NLP tasks. These models are pre-trained on large-scale unlabelled corpus and then usually fine-tuned on downstream tasks. 
Recent studies~\cite{le2021log, li2020_swisslog} demonstrate that these pre-trained models can be applied to understand the semantic meanings of log messages, thus favouring many downstream log analytics tasks. In this paper, we choose RoBERTa~\cite{liu2019roberta} as the studied pre-trained model since it is one of the most widely-used models. RoBERTa is an encoder-only model and uses the same transformer architecture as BERT~\cite{devlin2018bert}.  Different from BERT, RoBERTa is trained to predict the mask token with a large byte-level Byte-Pair Encoding (BPE)~\cite{sennrich2015neural}. One of the main reasons we choose RoBERTa over BERT is that the use of BPE allows RoBERTa to tokenize any input text without introducing any ``unknown" tokens \rv{by tokenizing out-of-vocabulary words into subwords}. This makes RoBERTa more suitable for log parsing because parameters created by developers are far beyond the scale of common English words and constantly changing, which would incur the out-of-vocabulary problem~\cite{liu2022uniparser}. Several studies also found that RoBERTa is effective for log analysis~\cite{le2021log, li2020_swisslog, guo2021logbert}.

\subsubsection{Virtual Label Token Generation}


Given an input sequence, $X = \{x_1, x_2, ..., x_n\}$, we adopt the 
template-free prompt tuning method~\cite{ma-etal-2022-template} to 
predict a virtual label token \logtext{PARAM} at the position $i$ via the pre-trained language model $\mathcal{M}$, where $x_i$ is a parameter.
Since all parameters are converted to the same token, it is essential to find a pivot token that can properly represent the parameters.

From the training set $\mathcal{D}_{train}\!=\!\{(X_i,\!Y_i) \}_{i=1}^K$, we leverage the pretrained language model $\mathcal{M}$ to get the probability distribution of predicting each token $t$ at each position $i$.
Specifically, we feed each sample $(X, Y)$ into $\mathcal{M}$ and get the probability distribution $p(\widehat{x}_i\,=\,t | X)$ of predicting each token $t$ in the log message $X$. Then, for each position $i$ which is indicated as a parameter, we select the top$k$ predicted tokens of $x_i$ as the initial parameters indication set $\mathcal{V}_{ini}$. This step aims to select top$k$ tokens having a similar meaning to the original parameter tokens to enrich the parameters indication set.


From the initial label-words set $\mathcal{V}_{ini}$, we simply search for the most frequent word in the unlabelled data. Specifically, we calculate the frequency $\phi(x = t | D)$ of each token $t \in \mathcal{V}_{ini}$ and select the most frequent words by ranking:
\begin{equation}
    \mathcal{V} = \underset{t}{\text{argmax}}~\phi(x = t | D), \forall t \in \mathcal{V}_{ini}
\end{equation}

After obtaining the set $\mathcal{V}$, we assign the embedding vector for the virtual label token \logtext{PARAM} by calculating the mean vector of all tokens in $\mathcal{V}$ and add it to the language model $\mathcal{M}$. 
\subsubsection{Training}
Given the input log message $X=\{x_1, x_2, \dots, x_n\}$, we construct a target sequence $Y=\{y_1, y_2, \dots, y_n\}$ by replacing the parameter at the position $j$ with the virtual label token \logtext{PARAM}, and maintaining the original words at keyword positions using Equation~\ref{eq:prompt-function}.
Then, the LM model is trained to maximize the probability $P(Y | X)$ of the target sequence $Y$:
\begin{equation}
    \mathcal{L} = - \frac{1}{K} \sum_{K}^{i=1} \Biggl(\frac{1}{n} \sum_{j=1}^{n} log P(x_j = y_j | X_i) \Biggl)
\end{equation}
where $K$ is the number of labelled training samples.

Note that we reuse the whole pre-trained model 
during the tuning process. The entity-oriented objective is similar to the LM-based (i.e., mask token prediction) objective, which can reduce the gap between pre-training and fine-tuning, thus allowing our model to keep the knowledge learned by the pre-trained LM model.

\subsection{Online Parsing}
\label{sec:online-parsing}

During online parsing (inference), we directly feed the log messages into the trained model, which will first tokenize the input to a set of tokens and then predict their corresponding target tokens. If a token is predicted as \logtext{PARAM}, it will be integrated into the parameter list; otherwise, it will be kept in the log template. Finally, we follow~\cite{khan2022guidelines} to post-process log templates by replacing consecutive parameters with a single parameter.
Note that we only need a one-pass decoding process to parse a log message, which is efficient when scaling to a large volume of logs.

%% file: sections/ex_design.tex
\subsection{Research Questions}
We evaluate our approach by answering the following research questions (RQs):

\textbf{RQ1:} How effective is LogPPT?

\textbf{RQ2:} How efficient is LogPPT?

\textbf{RQ3:} How do different modules contribute to LogPPT?

\textbf{RQ4:} How does \tool perform with different 
tuning techniques?

\subsection{Datasets}
We conduct experiments based on datasets initially collected from the \textit{LogPai} benchmark~\cite{zhu2019tools,loghub2021}, which consists of log data of 16 different systems, including distributed systems, supercomputers, operating systems, mobile systems, server applications, and standalone software. To determine the ground truth log templates, Zhu et al.~\cite{zhu2019tools} randomly sampled 2,000 log messages for each dataset and manually labelled them. However, recent studies~\cite{khan2022guidelines, liu2022uniparser} point out that there are multiple errors from these original datasets. Therefore, Khan et al.~\cite{khan2022guidelines} applied some heuristic rules such as Double Space or User-defined String to fix incorrect templates in the original datasets. In this study, we use the corrected version of these 16 datasets from~\cite{khan2022guidelines} in our evaluation.

\subsection{Baselines}
We compare our proposed method with five state-of-the-art methods, including AEL~\cite{jiang2008abstracting}, LenMa~\cite{shima2016length}, Spell~\cite{du2016spell}, Drain~\cite{he2017drain}, and Logram~\cite{dai2020logram}. These approaches apply many techniques such as similarity-based clustering (i.e., LenMa), frequency-based mining (i.e., AEL and Logram), or heuristics-based searching (i.e., Drain and Spell).
We choose these five approaches in our evaluation since they have their source code publicly available; and a prior study~\cite{zhu2019tools} finds that these approaches have the highest accuracy and efficiency among all the evaluated log parsers. We adopt the implementation of these methods from their replication packages~\cite{logparser2022, guideline2022}.

For a fair comparison, we extend baseline methods to include the labelled data from the data sampling phase. We transform the message-level labels into token-level labels by splitting log messages using default separators of each method.

\subsection{Evaluation Metrics}
Following recent studies~\cite{zhu2019tools, khan2022guidelines, liu2022uniparser, nedelkoski2020self-parsing}, we apply three metrics in our evaluation, including:

\textbf{Group Accuracy (GA)}: Group Accuracy ~\cite{zhu2019tools} is the most commonly used metric for log parsing. Group Accuracy considers template identification as a clustering process in which log messages with different log events are clustered into different groups~\cite{khan2022guidelines}. The GA metric is defined as the ratio of ``correctly parsed" log messages over the total number of log messages, where a log message is considered ``correctly parsed" if and only if it is grouped with other log messages consistent with the ground truth. However, recent studies~\cite{khan2022guidelines, liu2022uniparser} show that GA only accounts for how the parsed templates support the log message grouping activity instead of considering whether the templates and parameters are correctly identified or not. 


\textbf{Parsing Accuracy (PA):} The Parsing Accuracy (or Message-Level Accuracy~\cite{liu2022uniparser}) metric is defined as the ratio of ``correctly parsed" log messages over the total number of log messages, where a log message is considered to be ``correctly parsed" if and only if every token of the log message is correctly identified as template or variable. This metric is much stricter than Group Accuracy since any incorrectly parsed token will lead to the wrong parsing result for the whole log message.
We found that this metric is useful when evaluating the performance of log parsers when dealing with unseen log events compared to Group Accuracy. For example, for those log events that only appear once, GA always considers them as correctly identified since they belong to the correct groups. In contrast, PA could mark this identification as incorrect if some variables are incorrectly recognised as keywords.

\textbf{Edit Distance (ED):} Edit Distance is proposed in~\cite{nedelkoski2020self-parsing}. Different from GA and PA, Edit Distance is used to evaluate the template extraction in terms of string comparison. Specifically, Edit Distance (or Levenshtein edit distance) is computed by counting the minimum number of operations required to transform one template into the other~\cite{nedelkoski2020self-parsing}.
The score of Edit Distance for a dataset is computed as the median edit distance of all parsed template and ground truth template pairs.
By computing the distance between parsed templates and ground truth templates, this metric can measure the accuracy of log parsers in terms of meaning similarity (i.e., lexical similarity in our evaluation) between parsed results and ground truth. Note that the smaller the distance between two templates, the more similarity between them.
\subsection{Implementation and Environment}
\label{sec:env}
We conduct our experiments on a GPU server equipped with NVIDIA Tesla V100 GPU and CUDA 10.2. We implement \tool with Python 3.8 and PyTorch 1.7.
Followed recent studies for prompt tuning~\cite{wang2022no, gao2021making},
during the training process, we utilize AdamW~\cite{loshchilov2017decoupled} optimizer and set the initial learning rate to $5e^{-5}$. We set the batch size as 8 and train the model for 200 steps. AdamW optimizer is used with a linear decaying schedule with 10\% warm-up steps. During the online parsing phase, we set the batch size to 32.
In the Virtual Label Token Generation module, we calculate the embedding of the virtual label token \logtext{PARAM} from the 8 most frequent label tokens for our experiments.
We also evaluate the performance of \tool with different numbers of frequent label tokens. We provide the results in our project webpage\footnote{\repourl} due to space constraints. The results show that the performance of the proposed method is robust to the number of label tokens. It achieve consistently good results when choosing at least four label tokens.
In the Few-shot Data Sampling module, we set $K = 32$ as the default. We also experiment with different values of $K$ (from 4 to 128) in the experiments. 

%% file: sections/ex_results.tex
\subsection{RQ1: Parsing Effectiveness}
\label{sec:RQ1}
\subsubsection{Accuracy}
\begin{table*}[!htbp]
\caption{Comparison with the state-of-the-art log parsers with 32-shot ($\uparrow$: higher is better; $\downarrow$: lower is better)}
\label{tab:rq1_32shot}
\centering
\resizebox{\textwidth}{!}{
\setlength{\tabcolsep}{2.5pt}
\renewcommand{\arraystretch}{1.17}
\begin{tabular}{c|ccc|ccc|ccc|ccc|ccc|ccc} 
\toprule
\multirow{2}{*}{} & \multicolumn{3}{c|}{\textbf{AEL}}                     & \multicolumn{3}{c|}{\textbf{LenMa}}                   & \multicolumn{3}{c|}{\textbf{Spell}}                   & \multicolumn{3}{c|}{\textbf{Drain}}                   & \multicolumn{3}{c|}{\textbf{Logram}}                  & \multicolumn{3}{c}{\textbf{LogPPT}}                   \\ 
\cline{2-19}
                  & GA ($\uparrow$) & PA ($\uparrow$) & ED ($\downarrow$) & GA ($\uparrow$) & PA ($\uparrow$) & ED ($\downarrow$) & GA ($\uparrow$) & PA ($\uparrow$) & ED ($\downarrow$) & GA ($\uparrow$) & PA ($\uparrow$) & ED ($\downarrow$) & GA ($\uparrow$) & PA ($\uparrow$) & ED ($\downarrow$) & GA ($\uparrow$) & PA ($\uparrow$) & ED ($\downarrow$)  \\ 
\hline
HDFS              & 0.626           & 0.630           & 0.926             & 0.998           & 0.125           & 3.949             & \textbf{1}      & 0.487           & 0.858             & 0.998           & \textbf{0.959}  & 0.452             & 0.012           & 0.018           & 18.665            & \textbf{1}      & 0.902           & \textbf{0.276}     \\
Hadoop            & 0.677           & 0.422           & 12.163            & 0.667           & 0.242           & 16.788            & 0.533           & 0.196           & 9.274             & 0.948           & 0.439           & 7.564             & 0.283           & 0.370           & 19.014            & \textbf{0.994}  & \textbf{0.895}  & \textbf{0.882}     \\
Spark             & 0.415           & 0.381           & 3.088             & 0.869           & 0.023           & 10.130            & 0.920           & 0.336           & 4.192             & 0.905           & 0.376           & 2.568             & 0.282           & 0.275           & 7.433             & \textbf{0.999}  & \textbf{0.991}  & \textbf{0.167}     \\
Zookeeper         & 0.657           & 0.527           & 2.430             & 0.894           & 0.457           & 4.472             & 0.987           & 0.453           & 2.450             & 0.967           & 0.498           & 2.304             & 0.724           & 0.516           & 3.928             & \textbf{0.994}  & \textbf{0.990}  & \textbf{0.338}     \\
BGL               & 0.491           & 0.410           & 4.288             & 0.316           & 0.154           & 7.929             & 0.850           & 0.329           & 5.952             & 0.955           & 0.444           & 3.958             & 0.218           & 0.170           & 8.954             & 0.954  & \textbf{0.970}  & \textbf{0.233}     \\
HPC               & 0.731           & 0.698           & 1.151             & 0.681           & 0.671           & 2.687             & 0.657           & 0.532           & 3.633             & 0.741           & 0.672           & 1.845             & 0.742           & 0.679           & 2.628             & \textbf{0.943}  & \textbf{0.947}  & \textbf{1.147}     \\
Thunderbird       & 0.650           & 0.203           & 13.640            & 0.943           & 0.171           & 7.924             & 0.856           & 0.039           & 12.280            & \textbf{0.960}  & 0.191           & 13.675            & 0.129           & 0.128           & 15.479            & 0.679           & \textbf{0.926}  & \textbf{0.857}     \\
Windows           & 0.685           & 0.389           & 10.475            & 0.287           & 0.266           & 19.132            & 0.990           & 0.004           & 2.961             & \textbf{0.994}  & 0.696           & 4.705             & 0.694           & 0.374           & 6.413             & 0.991           & \textbf{0.983}  & \textbf{0.461}     \\
Linux             & 0.404           & 0.239           & 15.200            & 0.238           & 0.132           & 12.631            & 0.162           & 0.109           & 16.069            & 0.422           & 0.194           & 15.438            & 0.201           & 0.185           & 16.514            & \textbf{0.934}  & \textbf{0.949}  & \textbf{0.279}     \\
Android           & 0.642           & 0.559           & 8.082             & 0.778           & 0.722           & 5.602             & 0.891           & 0.241           & 8.311             & 0.765           & 0.730           & 5.626             & 0.677           & 0.428           & 12.872            & \textbf{0.885}  & \textbf{0.767}  & \textbf{1.143}     \\
HealthApp         & 0.570           & 0.175           & 18.474            & 0.166           & 0.289           & 15.947            & 0.961           & 0.152           & 5.119             & 0.644           & 0.241           & 18.393            & 0.258           & 0.263           & 15.173            & \textbf{1}      & \textbf{0.789}  & \textbf{2.536}     \\
Apache            & 0.984           & 0.987           & 0.189             & 0.984           & 0.293           & 3.524             & 0.301           & 0.285           & 10.275            & \textbf{1}      & \textbf{1}      & \textbf{0}        & 0.297           & 0.509           & 1.658             & \textbf{1}      & 0.994           & 0.024              \\
Proxifier         & 0.495           & 0.506           & 9.980             & 0.495           & 0.506           & 9.168             & 0.527           & 0.478           & 6.457             & 0.527           & 0.527           & 9.982             & 0.016           & 0               & 27.118            & \textbf{1}      & \textbf{1}      & \textbf{0}         \\
OpenSSH           & 0.198           & 0.421           & 4.193             & 0.927           & 0.155           & 8.744             & 0.488           & 0.127           & 5.888             & \textbf{0.996}  & 0.534           & 3.539             & 0.343           & 0.482           & 4.654             & 0.628           & \textbf{0.976}  & \textbf{0.119}     \\
OpenStack         & 0.266           & 0.187           & 9.822             & 0.213           & 0.191           & 11.199            & 0.245           & 0               & 19.663            & 0.224           & 0.187           & 20.801            & 0.241           & 0.112           & 49.110            & \textbf{0.989}  & \textbf{0.907}  & \textbf{0.788}     \\
Mac               & 0.583           & 0.223           & 18.523            & 0.648           & 0.155           & 20.867            & 0.724           & 0.033           & 23.281            & 0.711           & 0.277           & 20.531            & 0.551           & 0.252           & 21.651            & \textbf{0.780}  & \textbf{0.673}  & \textbf{8.856}    \\ 
\hline
Average           & 0.567           & 0.435           & 8.289             & 0.631           & 0.284           & 10.043            & 0.693           & 0.237           & 8.541             & 0.797           & 0.498           & 8.211             & 0.354           & 0.297           & 14.454            & \textbf{0.923}  & \textbf{0.916}  & \textbf{1.130}     \\
\bottomrule
\end{tabular}
}
\end{table*}

In this RQ, we compare LogPPT with five state-of-the-art methods (including AEL~\cite{jiang2008abstracting}, LenMa~\cite{shima2016length}, Spell~\cite{du2016spell}, Drain~\cite{he2017drain}, and Logram~\cite{dai2020logram}) on all 16 log datasets. 
Firstly, we compare the results of \tool with baselines using $K=32$ labelled samples.
The results in terms of three metrics (Group Accuracy, Parsing Accuracy, and Edit Distance) are shown in Table~\ref{tab:rq1_32shot}.

From the results, we can see that our model outperforms baseline methods on almost all datasets in the three evaluation metrics. Specifically, in terms of Group Accuracy (GA), \tool exceeds the most powerful log parser (Drain) by 15.8\% (0.923 versus 0.797 on average) and achieves the best results on 12 out of 16 datasets. It is worth noting that \tool achieves the accuracy of over 0.9 on 12 datasets and achieves 1.0 accuracy on four datasets among them, which is significantly superior to existing log parsers. In terms of Parsing Accuracy (PA), \tool surpasses baselines by at least 83.9\% when achieving an accuracy of 0.916 on average. \tool also achieves the best parsing accuracy on 14 out of 16 datasets. The high Parsing Accuracy suggests that \tool is able to accurately recognise the templates and corresponding parameters of log messages.
The experimental results 
confirm that \tool is effective in grouping logs into the same templates and identifying correct log templates and parameters.

Inspired by recent studies~\cite{khan2022guidelines, nedelkoski2020self-parsing}, we also evaluate our proposed \tool in terms of Edit Distance (ED) to measure the similarity between identified templates and their corresponding ground truth. It can be seen that \tool achieves the best average edit distance of 1.130, which is 7 times better than Drain. Besides, \tool outperforms baseline approaches on 15 out of 16 datasets and achieves a comparable result on the Apache dataset (0.024 versus 0). The experimental results on Edit Distance show that
the parsed templates produced by \tool have high textual similarities with the ground truth.
The main reason for the high accuracy of \tool is that it is capable of learning from semantic information of log messages, thus is able to precisely identify the templates and parameters of log messages.

\subsubsection{Robustness}

\begin{figure}[h]
    \centering
    \includegraphics[width=\linewidth]{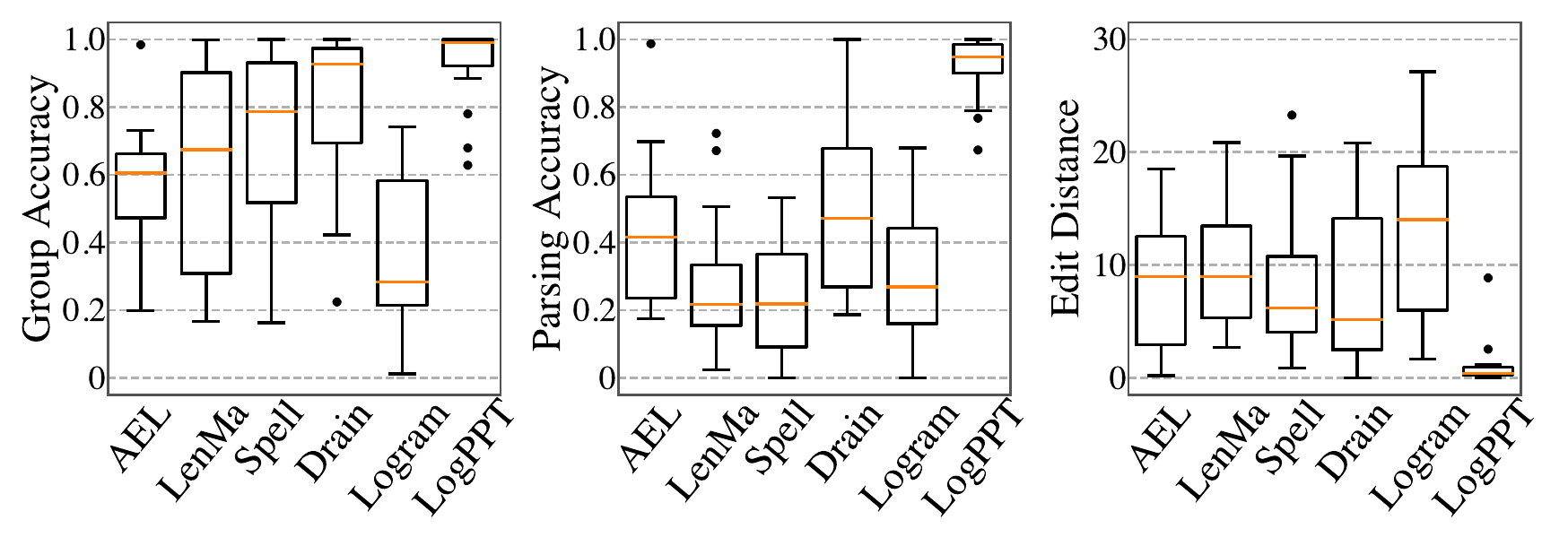}
    \caption{Accuracy Distribution of Log Parsers with 32-shot}
    \label{fig:rq1_acc_distribution}
\end{figure}

Our proposed \tool explicitly aims at supporting a broad range of diverse log datasets as employing a general log parser in production environments requires a robust performance~\cite{zhu2019tools}. Existing log parsers are sensitive to 
pre-processing steps, which involve domain-specific knowledge. Therefore, they show low robustness against different logging formats and behaviours~\cite{zhu2019tools, he2016evaluation}.
Therefore, we next analyze and compare the robustness against different types of logs of \tool with that of the baselines. Figure~\ref{fig:rq1_acc_distribution} shows the accuracy distribution of each log parser across different log datasets.

From the results, we can see that \tool outperforms the baselines in terms of robustness \rv{across different log types}. Existing methods require different regular expressions for pre-processing and different hyper-parameter values, thus, performing inconsistently on different datasets. For example, Drain uses different \textit{similarity threshold} (e.g., 0.2 for HealthApp and 0.6 for Proxifier) and different regular expressions (e.g., \logtext{blk\_-?$\backslash$d+} for HDFS and \logtext{core.$\backslash$d+} for BGL) for different datasets.
In contrast, \tool does not require to manually define regular expressions and achieves the smallest variance over different datasets. \tool is robust and performs well on most of the datasets (accuracy higher than 0.9) in terms of group and parsing accuracy. For example, \tool yields a median of 0.99 for GA robustness and 0.94 for PA robustness, which exceeds the second best log parser (i.e., Drain) by 6.9\%, and 98.5\%, respectively. Besides, \tool uses the same set of hyper-parameter values for every dataset in the training phase 
and does not require re-adjustment for each dataset.
Overall, the experimental results confirm that \tool is robust and can be applied to different log datasets with low effort.

\begin{figure}[ht]
    \centering
    \includegraphics[width=.9\linewidth]{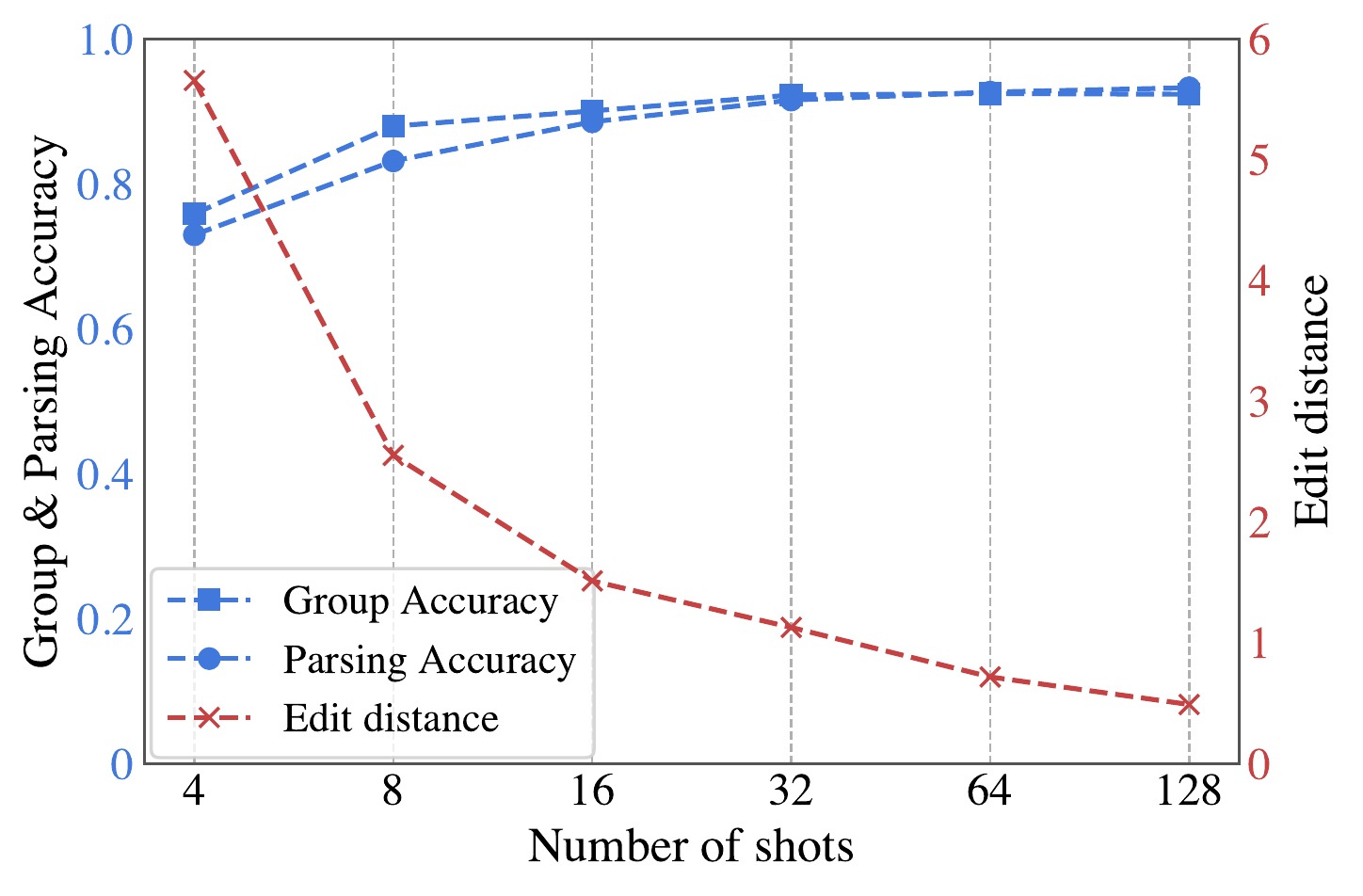}
    \caption{Results of LogPPT with different shots ($K$)}
    \label{fig:rq1_diff_shots}
\end{figure}

Our method requires a small amount ($K$) of labelled data sampled by an adaptive random sampling algorithm as the training set. Therefore, to evaluate the sensitivity of our proposed \tool to the amount of labelled data, we conduct an experiment using different numbers of training log messages (i.e., different \textit{shots}). Figure~\ref{fig:rq1_diff_shots} shows the performance of \tool with different numbers of shots. 

The experimental results show that the model's performance witnesses a severe drop when less data is used for training. The low results are reasonable since pre-trained models require task-specific data for better adapting to downstream tasks~\cite{wang2022no}. However, we observe that \tool achieves a good balance between Group Accuracy and Parsing Accuracy. \rv{Also,} \tool performs better than baselines in terms of Parsing Accuracy and Edit Distance even with only four labelled training samples. Moreover, it is noticeable that \tool can consistently achieve good results when $K \geq 16$.

In summary, \tool significantly outperforms the existing approaches in all three evaluation metrics. The experimental results confirm that \tool is capable of 
recognising log templates and the corresponding parameters.

\subsubsection{Accuracy with Unseen Logs}
Unseen log events occur frequently in logs. In this study, we consider the log events appearing only once in a dataset as previously unseen log events.
\tool can accurately recognise the templates and corresponding parameters of unseen log events, as reflected by the high Parsing Accuracy. To further evaluate the ability of \tool in parsing unseen logs, we measure the Parsing Accuracy of \tool on unseen log data and compare it with baseline methods. Specifically, for every dataset, we extract those log messages whose corresponding log templates only appear one time based on the ground truth, then calculate the Parsing Accuracy on these log messages.
Table~\ref{tab:res-unseen-logs} shows the results.
There are 42.64 unseen log events on average on 16 studied datasets. \tool achieves the best accuracy of 0.599 when parsing unseen log data, which exceeds existing log parsers by 58.9\% (LenMa) to 517.5\% (Logram).
\begin{table}[h]
\centering
\caption{Parsing accuracy on unseen log data}
\label{tab:res-unseen-logs}
\resizebox{\linewidth}{!}{
\setlength{\tabcolsep}{2pt}
\renewcommand{\arraystretch}{1.1}
\begin{tabular}{cccccccc} 
\toprule
                 & \#Unseen & AEL   & LenMa & Spell & Drain & Logram & LogPPT          \\ 
\hline
Parsing Acc. & 42.64         & 0.335 & 0.377 & 0.230 & 0.372 & 0.097  & \textbf{0.599}  \\
\bottomrule
\end{tabular}
}
\end{table}

\subsection{RQ2: Runtime Performance Evaluation}

\begin{figure}[h]
    \centering
    \includegraphics[width=.85\linewidth]{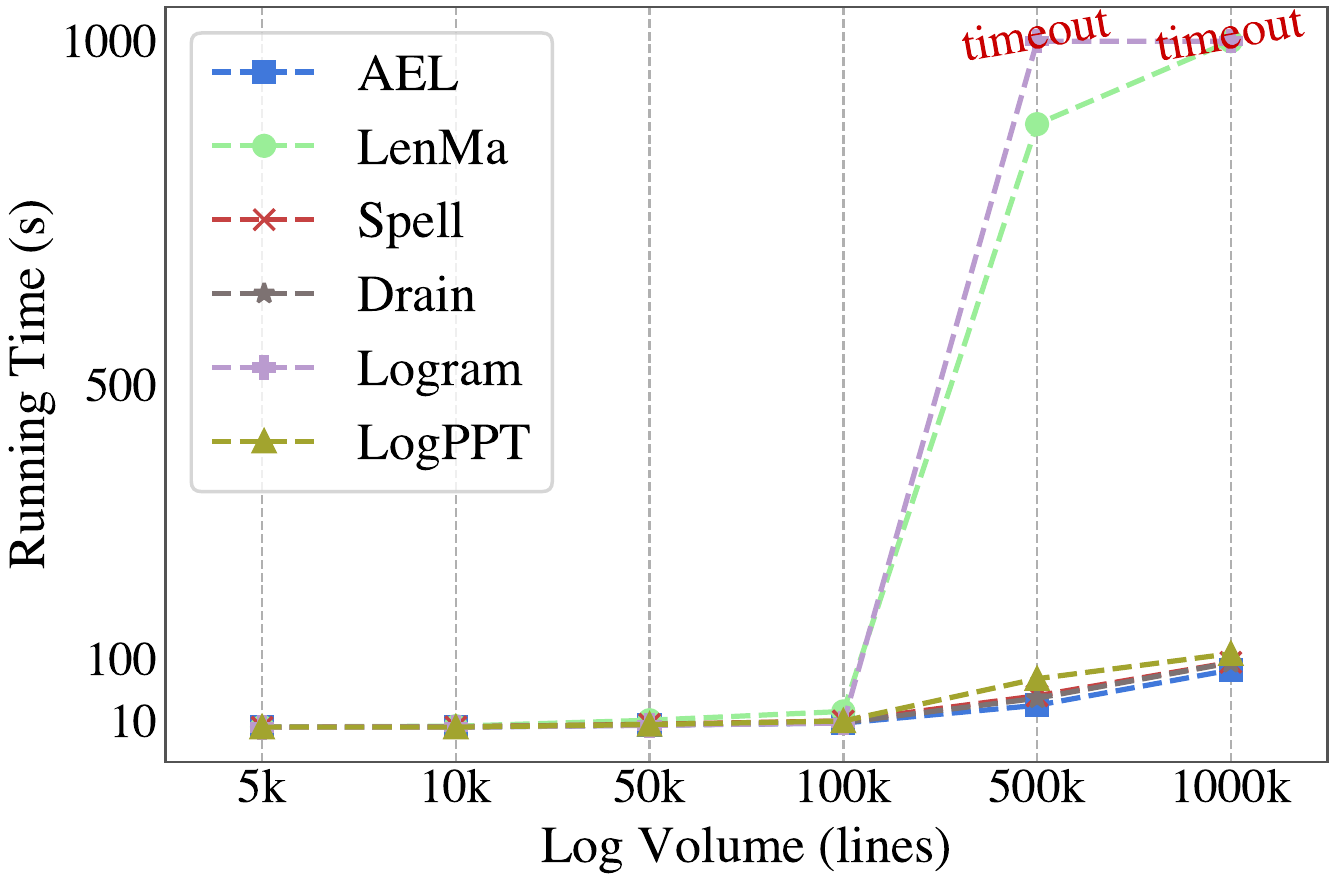}
    \caption{Running time of different log parsers under different volume}
    \label{fig:rq2_running_time}
\end{figure}

Besides effectiveness, efficiency is another critical metric for log parsers to consider in order to handle large-scale log data. To measure the efficiency of our proposed LogPPT, we record the running time it needs to finish the entire parsing process and compare it with the baseline methods. Specifically, we conduct this experiment on BGL and HDFS datasets, as they are relatively large. Figure~\ref{fig:rq2_running_time} reports the results.

We can see that the running time of \tool increases slowly with the increase of log data volume. With the use of GPU acceleration, our model can perform faster than or comparable with traditional log parsers. For example, \tool takes about 107 seconds to process one million log messages,
which is just slightly slower than Drain (94s), Spell (95s) and AEL (84s) and much faster than LenMa and Logram (cannot finish within 1,000 seconds). 

\subsection{RQ3: Ablation Study}

In this section, we evaluate the effectiveness of the major components and parameters in our proposed model.
Specifically, we exclude the Virtual Label Token Generation module and let the pre-trained model automatically assign the embedding for the virtual label token \logtext{PARAM}. To measure the contribution of the Adaptive Random Sampling module, we remove it from our model and randomly sample the log messages for labelling. We repeat this random process five times to avoid random bias and report the average results in Table~\ref{tab:ablation-study}. 

\begin{table}[h]
\centering
\caption{Ablation study results}
\label{tab:ablation-study}
\resizebox{.93\linewidth}{!}{
\setlength{\tabcolsep}{1.5pt}
\renewcommand{\arraystretch}{1.1}
\begin{tabular}{lccc} 
\toprule
                             & \textbf{GA} & \textbf{PA} & \textbf{ED}  \\
\hline
Full LogPPT                  & 0.923                   & 0.916                     & 1.130                   \\
w/o\textsubscript{Virtual Label Token Gen.}     & 0.879\textsubscript{($\downarrow$5.8\%)}                   & 0.835\textsubscript{($\downarrow$8.8\%)}                    & 3.130\textsubscript{($\downarrow$177\%)}                  \\
w/o\textsubscript{Adaptive Random Sampling} & 0.890\textsubscript{($\downarrow$3.6\%)}                  & 0.704\textsubscript{($\downarrow$23.1\%)}                    & 3.602\textsubscript{($\downarrow$219\%)}                  \\
\bottomrule
\end{tabular}
}
\end{table}

We can see that \tool performs worse in terms of parsing accuracy and edit distance without Virtual Label Token Generation and Adaptive Random Sampling modules. For example, without the Virtual Label Token Generation module, \tool only achieves a parsing accuracy of 0.835, which is 8.8\% worse than complete LogPPT, while it can still achieve an acceptable group accuracy (0.879). The reason is that without the Virtual Label Token Generation module, the model cannot find the pivot word that can mostly represent parameters in log messages. Consequently, many parameters are misidentified, leading to a worse parsing accuracy and edit distance. On the other hand, log messages are highly imbalanced under different log templates. Using a naive random sampling technique cannot guarantee the quality of the training set. Therefore, the results significantly decline when we remove the Adaptive Random Sampling module (23.1\% decreasing in terms of Parsing Accuracy). 

In summary, this comparison demonstrates the usefulness of the proposed Adaptive Random Sampling module and the Virtual Label Token Generation module of LogPPT.






\subsection{RQ4: Comparison with Different Tuning Techniques}
\tool applies a novel prompt tuning method (i.e., template-free prompt~\cite{ma-etal-2022-template}), which relaxes the burden of manually selecting prompt templates and improves the efficiency compared to other prompt tuning methods. In this section, we evaluate the performance of this prompt tuning method. To this end, we replace our prompt tuning module with four different prompt tuning methods (introduced in Section~\ref{sec:language-models}) and a fine-tuning technique. We then compare the performance of \tool with that of the variants.
\begin{itemize}
    \item \textbf{FT} (fine-tuning): We add a binary classification layer on top of the pre-trained RoBERTa model and fine-tune the model to perform log parsing as a binary token classification problem.
    \item \textbf{HardPT$_\text{M}$} (hard prompt tuning with manual label words): We use a standard hard prompt~\cite{cui2021template} with the prompt template of ``[X] [S] is a [MASK]", where [X], [S], and [MASK] are the unfilled slots for the input log message, token, and label respectively. The model learns to predict the label word at the [MASK] position. In this setting, we use fixed manual sets of label words, including ``\textit{[const, keyword]}" for keyword tokens and ``\textit{[variable, parameter]}" for parameter tokens.
    \item \textbf{HardPT$_\text{S}$} (hard prompt tuning with soft label words): We use the same standard hard prompt template as the above setting. However, we use trainable tokens~\cite{hambardzumyan2021warp} as the label words for this setting.
    \item \textbf{SoftPT$_\text{M}$} (soft prompt tuning with manual label words): In this setting, we follow recent works to use a soft prompt-template of ``[X] [S] [SOFT] [SOFT] [SOFT] [MASK]", where [X], [S], and [MASK] are the unfilled slots for the input log message, token, and label respectively. [SOFT] is a trainable token. The embeddings of these [SOFT] tokens are optimized during the tuning stage. We use manual label word sets for this setting as in \textbf{HardPT$_\text{M}$}.
    \item \textbf{SoftPT$_\text{S}$} (soft prompt tuning with soft label words): We use the same soft prompt template of ``[X] [S] [SOFT] [SOFT] [SOFT] [MASK]" as in the \textbf{SoftPT$_\text{M}$} setting. For label words, we adopt the same \textbf{HardPT$_\text{M}$} setting to use trainable tokens~\cite{hambardzumyan2021warp} as the label words.
\end{itemize}

\begin{table}[h]
\centering
\caption{Comparison with different tuning methods}
\label{tab:diff-tuning}
\resizebox{.95\linewidth}{!}{
\setlength{\tabcolsep}{2pt}
\renewcommand{\arraystretch}{1.1}
\begin{tabular}{c|ccc|ccc|ccc|ccc} 
\toprule
                & \multicolumn{3}{c|}{4shot}                    & \multicolumn{3}{c|}{8shot}                    & \multicolumn{3}{c|}{16shot}                   & \multicolumn{3}{c}{32shot}                    \\ 
\cline{2-13}
                & GA            & PA            & ED            & GA            & PA            & ED            & GA            & PA            & ED            & GA            & PA            & ED             \\ 
\hline
FT     & 0.69          & 0.70          & 5.98          & 0.88          & 0.74          & 3.56          & 0.85          & 0.84          & 2.09          & 0.91          & 0.91          & 1.23           \\
HardPT$_\text{M}$     & 0.54          & 0.54          & 6.75          & 0.64          & 0.62          & 4.12          & 0.69          & 0.64          & 3.57          & 0.68          & 0.68          & 2.71           \\
HardPT$_\text{S}$ & 0.48          & 0.44          & 10.08         & 0.57          & 0.51          & 8.83          & 0.54          & 0.51          & 8.47          & 0.67          & 0.67          & 3.27           \\
SoftPT$_\text{M}$     & 0.56          & 0.52          & 6.63          & 0.54          & 0.61          & 5.05          & 0.54          & 0.55          & 8.26          & 0.66          & 0.65          & 5.38           \\
SoftPT$_\text{S}$     & 0.29          & 0.24          & 20.34          & 42          & 0.46          & 9.40          & 0.46          & 0.48          & 7.30          & 0.58          & 0.64          & 5.94           \\
LogPPT          & \textbf{0.76} & \textbf{0.73} & \textbf{5.66} & \textbf{0.88} & \textbf{0.83} & \textbf{2.56} & \textbf{0.90} & \textbf{0.89} & \textbf{1.51} & \textbf{0.92} & \textbf{0.92} & \textbf{1.13}  \\
\bottomrule
\end{tabular}
}
\end{table}

Table~\ref{tab:diff-tuning} shows the results. We can see that \tool with our proposed prompt tuning method achieves the best results among all studied methods. For example, with \textit{16shot} setting, \tool outperforms others by 6.0\%-74.5\% in terms of Parsing Accuracy. Our proposed method significantly outperforms other prompt tuning methods because it can leverage both semantic and position information of tokens in log messages. Standard prompt tuning methods \rv{overly} focus on leveraging the semantic meaning of a token and overlook the contextual information which is important in log parsing. Fine-tuning, on the other hand, can achieve better results than standard prompt tuning because it can use the positional information during the training stage. With more labelled training data, fine-tuning can achieve quite similar results with LogPPT.

\begin{figure}[h]
    \centering
    \includegraphics[width=.85\linewidth]{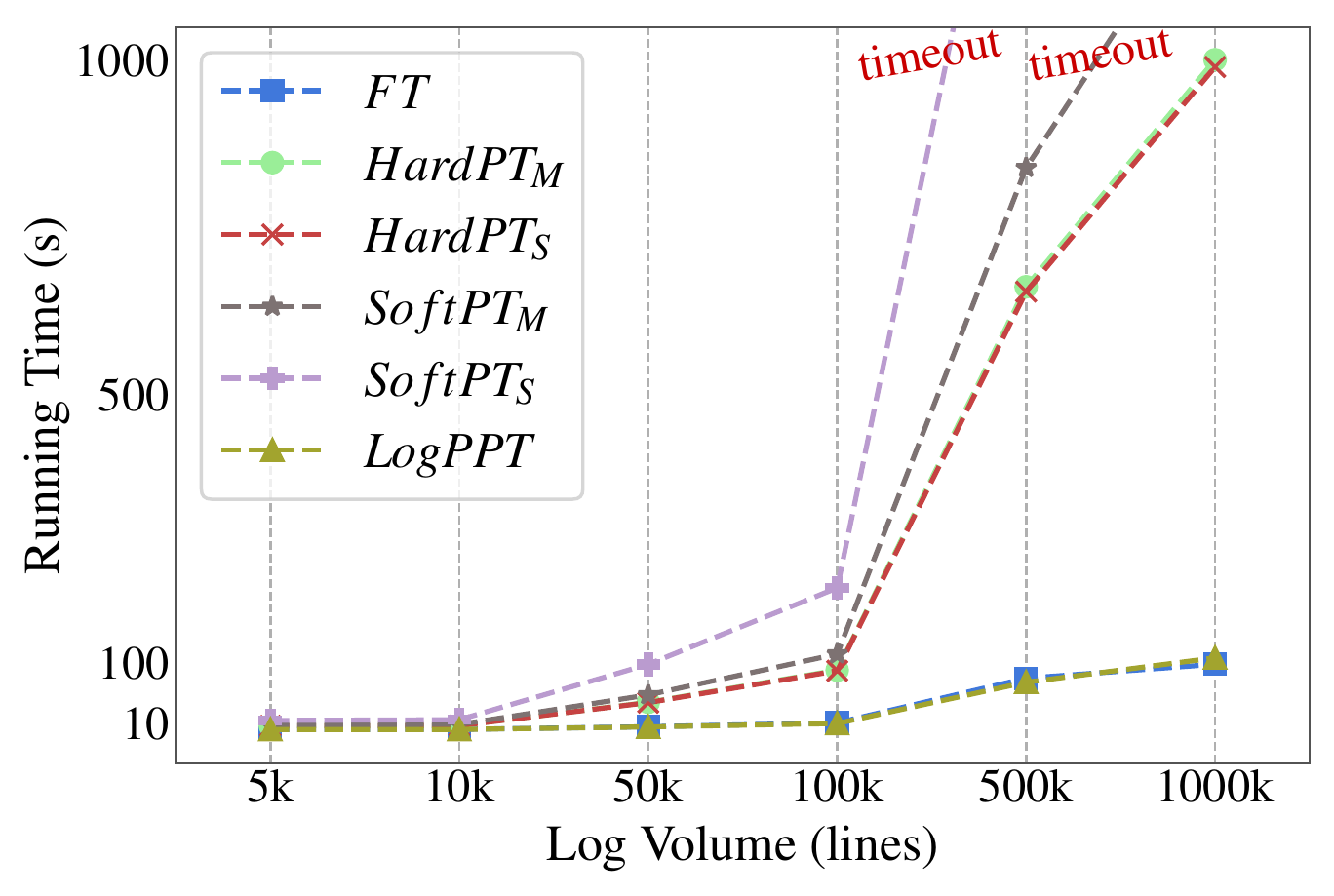}
    \caption{Parsing time of different tuning methods}
    \label{fig:rq3-parsing-time}
\end{figure}

Next, we evaluate the parsing time of different tuning methods. As shown in Figure~\ref{fig:rq3-parsing-time}, the parsing time of \tool and fine-tuning approach is similar because they only need one-pass decoding to parse one log message. On the other hand, other prompt tuning methods need to enumerate all tokens in a log message which is a time-consuming process. For example, with soft prompts, the model cannot finish parsing one million log lines within 1,000 seconds.

In summary, our proposed method is more effective and efficient compared to other tuning techniques and can achieve high accuracy with a few shots of training data.

%% file: sections/discussion.tex



\subsection{Why does \tool Work?}
\label{sec:diss-why-logppt-work}
There are several reasons that make \tool perform better than the related approaches. First, \tool predicts keywords and parameters using the semantic information from log messages by tuning a pre-trained language model. Thus, compared to traditional methods using only superficial features, \tool is able to indicate the keywords or parameters more precisely. Besides, \tool does not require domain-specific knowledge to define regular expressions 
for each dataset, thus is easy to be applied to a new log dataset.

Second, compared to other few-shot learning techniques, \tool applies an effective and efficient prompt tuning method, which can avoid the complex design for prompt instructions and also boost the few-shot performance. \tool leverages both semantic and positional information of tokens in log messages, thus can handle the noise in log data compared to other prompt tuning methods. For example, the log message from Proxifier, \logtext{open through \textbf{proxy} \textbf{proxy}.cse.cuhk.edu.hk:5070}, contains two \logtext{proxy} tokens with different roles. Standard prompt tuning methods fail to distinguish these tokens and predict the same label for them. The reason is that standard prompt tuning methods only consider the semantic meaning of tokens but ignore the position information, which is important for log parsing. In contrast, our method utilizes both semantic and position information of a token in log messages and achieves high parsing accuracy (100\% parsing accuracy).


\subsection{Threats to Validity}
\label{sec:threat-to-validity}
We have identified the following major threats to validity.

\textbf{Data Quality.} In this paper, we used public log datasets for our evaluation. The ground truth templates of all log messages, including log templates and corresponding parameters, are provided within the datasets. Although these datasets are commonly used by many related works~\cite{zhu2019tools, dai2020logram, nedelkoski2020self-parsing}, they may also contain a small proportion of errors. To reduce this threat, we leverage the latest version of the benchmark datasets~\cite{khan2022guidelines} that are corrected with automatic and manually-defined rules.

\textbf{Tool Comparison.} In our evaluation, we compared our results with related approaches. The approaches achieved the best results in a recent benchmark~\cite{zhu2019tools} and are used in both industry and academia. We adopt the implementations from their replication packages. We apply the parameters and settings (e.g., number of log templates, similarity threshold, etc.) optimized by the previous work~\cite{zhu2019tools}.

\textbf{Labelling Effort.} Our proposed method relies on a small number of labelled log data. To reduce the labelling effort, we propose to use an Adaptive Random Sampling algorithm to select a diverse set of $K$ log messages ($K$ from 4 to 128) and attain the templates from user feedbacks.

%% file: sections/related_work.tex
\textbf{Log Analysis with Language Models:}
Log analysis is a research area that has attracted lots of attention due to its practical importance. Typical applications of log analysis include anomaly detection~\cite{du2017deeplog, he2016experience,li2022swisslog, le2022log}, failure prediction~\cite{das2018desh, zhang2018prefix}, root cause analysis~\cite{lu2017log, gurumdimma2016crude}, etc. Recently, inspired by the success of pre-trained models in NLP, many studies have been proposed to apply pre-trained language models to log analysis. SwissLog~\cite{li2020_swisslog} and NeuralLog~\cite{le2021log} utilize the pre-trained BERT~\cite{devlin2018bert} model for log-based anomaly detection. Ott et al.~\cite{ott2021robust} studied the use of different pre-trained models such as BERT~\cite{devlin2018bert} and XLNet~\cite{yang2019xlnet} for log anomaly detection. Setianto et al.~\cite{setianto2021gpt} proposed to fine-tune the GPT2~\cite{radford2019language} model for log parsing.

\textbf{Data-driven Log Parsing:}
Log parsing has become an active research topic in recent years~\cite{he2017drain, du2016spell, nedelkoski2020self-parsing, ahmad2021unified}. 
Recently, to address the limitations of traditional log parsers and improve the parsing accuracy, some approaches~\cite{tao2021logstamp, liu2022uniparser} proposed to use token classification for log parsing. 
LogStamp~\cite{tao2021logstamp} converts the log parsing task into a sequence labelling problem. It leverages the BERT~\cite{devlin2018bert} model to classify words in log messages. These approaches, however, adopt a traditional log parser to generate pseudo labels for log messages as the training data, which can introduce many noises in training data. Liu et al.~\cite{liu2022uniparser} proposed UniParser, which is a unified log parser for heterogeneous log data. UniParser is trained with labelled data across multiple log sources to capture the common patterns of templates and parameters. 
Although effective, UniParser requires a noticeable amount of labelled data to train a classification model, which is not always available in practice. Besides, UniParser requires handcrafted rules to split raw log messages into tokens, which is not suitable to apply on some special dataset~\cite{liu2022uniparser}.

Our \tool can effectively leverage semantic information from a few labelled data by using a pre-trained language model. \tool does not require any domain-specific knowledge to pre-process log data, thus can adapt to new log dataset with low effort. Besides, by using a novel prompt tuning method, \tool can effectively learn the semantic patterns from a few labelled data. 

%% file: sections/conclusion.tex
Log parsing is the foundation step to enabling automated log analytics. To overcome the limitations of existing log parsers, we propose a log parser with prompt-based few-shot learning, namely LogPPT, to capture the patterns of templates and parameters. \tool utilises a novel prompt tuning method to recognise keywords and parameters from a few labelled log data selected by an adaptive random sampling algorithm. We have evaluated \tool on public log datasets. The results show that \tool is effective and efficient, outperforming the state-of-the-art log parsers. In the future, we will deploy \tool in a production environment to further evaluate its scalability and effectiveness in practice.

\textbf{Data Availability:} Our source code and experimental data are publicly available at \repourl.